\documentclass[%
reprint,nofootinbib,
 amsmath,amssymb,
 aps,
prb,
]{revtex4-2}

\usepackage{graphicx}
\usepackage{dcolumn}
\usepackage{bm}
\usepackage{braket}
\usepackage{amsmath}
\usepackage{amssymb}
\usepackage{pifont}

\begin{document}

\title{Wilson Line and Disorder Invariants of Topological One-Dimensional Multiband Models}

\author{R. Moola}
\author{A. Mckenna}
\author{M. Hilke}

\affiliation{%
 Department of Physics, McGill University, 3600 rue University, Montr\'eal, QC-H3Y 1H2, Canada
}%


\begin{abstract}
For simple models like the Su–Schrieffer–Heeger (SSH) model, the characterization of its topology via the winding number and Zak phase is straightforward; however, in multiband systems, this is no longer the case. In this work, we introduce the unwrapped Wilson line across the Brillouin zone to compute the bulk topological invariant. This method can be efficiently implemented numerically to evaluate multiband SSH-type models, including models that have a large number of distinct topological phases. This approach accurately captures all topological edge states, including those overlooked by traditional invariants, such as the winding number and Zak phase. To make a connection with the experiments, we determine the sign of the topological invariant by considering a Hall-like configuration. We further introduce different classes of disorder that leave certain edge states protected, while suppressing other edge states, depending on their symmetry properties. Our approach is illustrated using different one-dimensional models, providing a robust framework for understanding topological properties in one-dimensional systems.

\end{abstract}

\maketitle


\section{\label{sec:intro}Introduction} Topological invariants are quantized properties of gapped Hamiltonians that remain invariant under continuous deformations. They predict the presence of symmetry-protected edge states through the bulk-boundary correspondence principle \cite{jackiw1976solitons,ryu2002topological}. Well-known topological invariants, such as the winding number, Chern number, Zak phase, and polarization, characterize bulk properties and are specific to the symmetries and dimensionality of the system \cite{grossmann2016index}. Depending on the model, the class of invariant used follows from the 10-fold classification \cite{schnyder2008classification}. For the SSH model, the most famous of 1D topological insulators, a $\mathbb{Z}$ and $\mathbb{Z}_2$ invariant can be used to describe it's topology. The winding number ($\mathbb{Z}$) directly counts zero-energy chiral symmetry-protected edge states. The Zak phase, closely related to polarization ($\mathbb{Z}_2$), provides an alternative approach via the Wilson loop \cite{benalcazar2017quantized}. However, these invariants, have limitations; the winding number relies critically on chiral symmetry and can only count edge states pinned at zero energy, while the polarization, being a defined modulus a unit cell, cannot distinguish complex topological phases beyond $0$ and $1$ \cite{obana2019topological,kim2020topological,ma2022electronic}. To address these issues, we introduce a method that generalizes the Zak phase by unwrapping the Wilson line spectrum across the Brillouin zone (BZ). Unlike the Wilson loop, which is a closed path over the BZ, the Wilson line is an open-path object whose accumulated phase evolution encodes finer topological structure. By tracking the continuous phase, after unwrapping $2\pi$ discontinuities, we define a robust $\mathbb{Z}$ integer-valued topological invariant $\gamma$ applicable even when traditional invariants fail. 

This method is particularly useful for the multiband SSH$_4$ model defined in section III, which lacks inversion symmetry and features edge states at non-zero energy which are overlooked by the winding number and polarization \cite{bid2022topological,marques2020analytical,eliashvili2017edge,anastasiadis2022bulk}. Models with inversion symmetry have been discussed in detail in \cite{hughes2011inversion}, as well as non-zero energy edge states. Although these non-zero edge states within the band subgap of the SSH$_4$ do not have a gap closing to indicate a topological transition, works such as \cite{ryu2002topological} suggest that a change in the symmetry responsible for protecting said states can also signify a topological phase transition.  

In recent years, experimental studies have significantly improved our understanding of SSH models across multiple dimensions, with several studies reporting direct measurements of edge states in these systems \cite{geng2022observation,caceres2022experimental, zhang2025observation}. We further examine the robustness of the edge states captured by the invariant $\gamma$ under dilute disorder, highlighting its potential relevance for future experimental exploration.

\section{\label{sec:zak}Topological Invariants}

\subsection{\label{sec:wilson} Zak phase}

 The Zak phase is a topological invariant used to characterize one-dimensional (1D) periodic systems. It is defined as the geometric phase acquired by the electronic wavefunction as it traverses the entire Brillouin zone. It is equivalent to the Berry phase in 1D systems across a non-contractible loop \cite{zak1989berry}. The Zak phase has been shown to be related to polarization \cite{benalcazar2017quantized} and winding number \cite{Károly_László_András_2016}. The Zak phase is defined as follows:

\begin{equation}
\label{eq:one}
Z = \int_{-\pi}^{\pi} A(k) dk,
\end{equation}
where \(A(k) = i \braket{u(k) | \partial_k | u(k)}\) represents the  Berry connection of the Bloch eigenstate \(\ket{u(k)} \). This is well-defined for single-band systems but fails in multiband cases, where the eigenstates \(\ket{u_n(k)} \) will have band index $n$. In a multi-band system, the Berry connection $A(k)$ generalizes to a matrix, requiring a path-ordered integral rather than a simple line integral. The Wilson loop, which we introduce next, correctly tracks the evolution of the subspace of occupied bands across the BZ.
 
\subsection{\label{sec:wilson} Wilson Loop Polarization}

Polarization provides a fundamental tool for defining topological invariants in various crystalline insulators \cite{king1993theory, vanderbilt1993electric}. Polarization can be conveniently calculated numerically by first discretizing the Brillouin zone into N segments.
\begin{equation}
k_i = \frac{2\pi i}{N}, \quad i = 0, 1, \ldots, N-1.
\end{equation}
In a small segment from \(k_i\) to \(k_{i+1}\), the acquired phase \(\vartheta_i\) is given by 
\begin{equation}
\vartheta_i = \arg \braket{u_{k_i} | u_{k_{i+1}}}.
\end{equation}
Thus, the total Zak phase is the summation of the  phases of each segment:
\begin{equation}
\vartheta = \sum_{i=1}^{N} \vartheta_i.
\end{equation}
This can be expressed as
\begin{equation}
\vartheta = \arg \prod_{i=1}^{N} \braket{u_{k_i} | u_{k_{i+1}}}.
\end{equation}

This product defines the Wilson loop, a gauge-invariant measure computed over the Brillouin zone where $\ket{u(k_{0})} = \ket{u(k_N)}$. When considering multiple energy bands, the Berry phase for each segment is represented by the matrix \(M_{k_i k_{i+1}}\):
\begin{equation}
(M_{k_i k_{i+1}})_{nm} = \braket{u_n(k_i) | u_{m}(k_{i+1})},
\end{equation}
where the indices \(n\) and \(m\) run over all occupied bands. If there are \(\mathcal{N}\) bands below the Fermi level, then \(M_{k_i k_{i+1}}\) is a \(\mathcal{N} \times \mathcal{N}\) matrix. We can then take the matrix product:
\begin{equation}
\mathcal{W} = \prod_{i=1}^{N} M_{k_i k_{i+1}}.
\end{equation}
The phases associated with the eigenvalues are given by:
\begin{equation}
\vartheta^{(n)} = -\Im \log(\lambda_n), \quad n = 1, \ldots, \mathcal{N},
\end{equation}
where \(\lambda_n\) is the \(n\)-th eigenvalue of the matrix \(\mathcal{W}\). The total Zak phase is the sum over the occupied bands.
\begin{equation}
\vartheta = \sum_{n=1}^\mathcal{N} \vartheta^{(n)}.
\end{equation}

This formalism is gauge-invariant, ensuring that arbitrary phase choices for the Bloch wavefunctions and gauge transformations do not affect the polarization. The eigenvalues of the Wilson loop correspond to Wannier centers, which represent the electronic charge centers within the unit cell. The phases of these eigenvalues are directly related to the polarization, reflecting the bulk charge distribution across the unit cell. Specifically, the polarization $\mathbb{P}$ defined through this method is related to the Zak phase by the expression $Z = 2\pi \mathbb{P}$. Consequently, the Zak phase calculated via the Wilson loop spectrum is only defined modulo $2\pi$, as polarization is a geometric property that is defined on a lattice. This modularity arises because shifting the Wannier centers by a lattice constant does not alter the net dipole moment in a periodic crystal, thus preserving the physical polarization.
 
\subsection{\label{sec:winding}Winding Number}

The winding number characterizes the topology of the Bloch Hamiltonian by quantifying the number of times its mapping of the Brillouin zone ($k \mapsto Q(k)$) encircles the origin in the complex plane. This invariant establishes a direct bulk-boundary correspondence, determining the number of topologically protected zero-energy edge states. Given a chiral symmetric Hamiltonian \(H(k)\), the chiral symmetry operator \(\Gamma\) satisfies:
\begin{equation}
\Gamma H(k) \Gamma^{-1} = -H(k).
\end{equation}

The winding number is gauge-invariant as it can be computed directly from the Bloch Hamiltonian independently of the choice of eigenstate gauge. In the eigenstate basis of the chiral symmetry operator $\Gamma$, the Hamiltonian can be written as off-diagonal 
\begin{equation}
H(k) = \begin{pmatrix}
0 & Q(k) \\
Q^\dagger(k) & 0
\end{pmatrix}.
\label{eq:offdiag}
\end{equation}

In 1D, the winding number \(\nu\) is defined using the off-diagonal elements \(Q(k)\) and can be expressed as \cite{note1}

\begin{equation}\label{eq:winding}
\nu = \frac{i}{2\pi} \int_{-\pi}^{\pi} \text{Tr} \left( Q^{-1}(k) \frac{\partial Q(k)}{\partial k} \right) \, dk.
\end{equation}
$Q$ can be a number or a matrix. A drawback of this method is that it ceases to be well-defined when chiral symmetry is broken and fails in the bulk-boundary correspondence for non-zero energy edge states in models such as SSH$_4$ as we will see below. The winding number can be related to the Wilson loop Zak phase as follows: \cite{Károly_László_András_2016}
\begin{equation}
Z = \nu \pi(\text{mod}  2\pi).
\end{equation}

\subsection{$\mathbb{Z}$ Wilson Line Invariant, $\gamma$}
The concept of Wilson lines originates from gauge theory and was introduced by Wilson in the context of lattice gauge theory to study confinement in QCD \cite{wilson1974confinement}. While initially developed in high-energy physics, the closed-loop counterpart of a Wilson line, the Wilson loop, has found significant applications in condensed matter physics. Beyond computing Berry phases and Chern numbers, Wilson loops have been used to detect fragile topology \cite{bouhon2019wilson}, diagnose topological obstructions via Wannier charge evolution \cite{alexandradinata2014wilson}, and provide a geometric framework for nonlinear optical responses in topological materials \cite{zhou2022generalized}. While the gauge invariant Wilson loop method cannot be used to compute a \(\mathbb{Z}\) topological invariant for non-inversion symmetric multiband models directly, the behaviour of the Wilson line spectrum across the BZ can reveal such an invariant. We define the Wilson line as follows.

\begin{equation}
\mathcal{W}(\kappa) = \prod_{k_j = -\pi}^{\kappa} M_{k_j, k_j + dk}, \quad dk = \frac{2\pi}{N}
\end{equation}

The phase of the eigenvalues $\mathcal{W}(\kappa)$ depends on the endpoint ($\kappa$) of the Wilson line: 
\begin{equation}
\vartheta^{(n)}(\kappa) = -\Im \log(\lambda_n(\kappa))
\end{equation}

As we vary $\kappa$ from $-\pi$ to $\pi$, we get a list of phases.

\begin{equation}
\{\vartheta^{(n)}(-\pi),\vartheta^{(n)}(-\pi+dk),\hdots 
\vartheta^{(n)}(\pi)\}
\end{equation}

The phase exhibits discontinuities with jumps of $2 \pi$. We reconstruct a smooth and continuous phase function to extract the full accumulated phase across the BZ zone. To this end, we implement the use of an unwrapping process. Phase unwrapping first detects discontinuities by looking at the difference between points, then removes them by adding or subtracting multiples of \(2 \pi \). This process is used in applications such as interferometry, where accurate phase information is necessary for precise measurements \cite{itoh1982analysis}. Before unwrapping, we sum the original phase over all occupied bands. The topological Wilson line invariant $\gamma$ can be calculated from the unwrapped (uw) continuous phase function $\vartheta^{\text{uw}}(\kappa)$ \cite{note2}

\begin{equation}
\gamma =\frac{1}{\pi}\int_{-\pi}^{\pi} \frac{\partial}{\partial \kappa} \vartheta^\text{uw}(\kappa) \, d\kappa = (\vartheta^\text{uw}(\pi) - \vartheta^\text{uw}(-\pi)).
\label{eq:gamma}
\end{equation}

 The unwrapping process results in $\gamma$ being some integer of $2 \pi$ beyond the Wilson Loop Zak phase. Therefore, whenever the Zak phase is quantized, that is, when chiral or inversion symmetry is present, $\gamma$ will also be quantized.  It is important to note that $\gamma$ is gauge dependent. We choose a gauge such that the first or last element of $u(k)$ is real. If taking a gauge where the last element is real, consistency is ensured by aligning the first site of the right semi-infinite chain with a $A$ site in real space, regardless of how many sites are present in the model. This gauge is equivalent to setting the virtual site $D_i = 0 $ where the site index $i = 0$, so that no electron can hop to the left of $A_1$. The topological invariant calculated using this gauge will give us the number of edge states on the left boundary of a right semi-infinite chain. If choosing the first element of the eigenstate to be real, the same logic applies, and $\gamma$ counts the number of right edge states of a left semi-infinite chain. The Wilson line invariant can be computed numerically very effectively, and we will use it to evaluate the topology of 1D models and confirm the bulk-boundary correspondence. We further study the robustness of the topological edge states to different types of disorder.

\section{\label{sec:1d lattice} 1D Lattice Models}

\begin{figure}[h]
        \includegraphics[width=\linewidth]{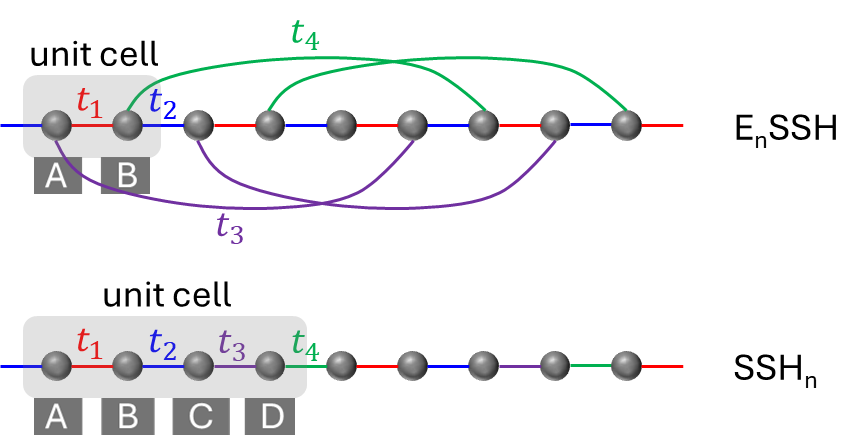}
        \caption{Top drawing illustrates the E$_n$SSH, which is an extended SSH model with two sites, A and B, per unit cell. In this example, the E$_2$SSH is plotted. The bottom drawing illustrates the SSH$_n$ model, where the unit cell is composed of $n$ atoms linearly coupled by $t_i$. In this example, the SSH$_4$ has a unit cell composed of A, B, C, and D sites.}
        \label{fig:ESSH}
\end{figure}

1D tight-binding models play an important role in understanding topological properties. We will consider several generalizations of the SSH model belonging to the BDI symmetry class. The SSH$_n$ model generalizes the SSH chain to a unit cell of size $n$, while the extended SSH model , E$_n$SSH, has a unit cell of two sites (A and B) but allows long-range hopping over a distance of $n$ unit cells, always between A and B sites to preserve chiral symmetry.These models are illustrated in fig. \ref{fig:ESSH}. We will start with the simplest topological model, the SSH model, which is equivalent to SSH$_2$ and E$_0$SSH in our notation. 

\subsection{Dimer SSH}

The SSH model has two sites (A and B) per unit cell with alternate hopping $t_1, t_2$.
The Bloch Hamiltonian for the SSH model can be written as
\begin{equation}
H(k) = \left( \begin{array}{cc}
0 & t_1 + t_2 e^{-ik} \\
t_1 + t_2 e^{ik} & 0
\end{array} \right).
\end{equation}

In the gauge, where the second element is set to be real and positive, the corresponding Bloch eigenstates can be written as

\begin{equation}
| u_\pm(k) \rangle = \frac{1}{\sqrt{2}} \left( \begin{array}{c}
\pm e^{i\varphi(k)} \\ 1
\end{array} \right) = \left( \begin{array}{c}
\text{A-site} \\ \text{B-site}
\end{array} \right),
\end{equation}
where \( \varphi(k)  =\arg(  t_1 + t_2 e^{-ik}) \). We will impose this gauge condition in the following analytical and numerical calculations. This gauge ensures consistency in the calculation of topological invariants by aligning the first site of the left semi-infinite chain with the $A_1$ site, regardless of how many sites are present in the model. This gauge is equivalent to setting the virtual site $D_0 =0 $, so that no electron can hop to the left of $A_1$. The topological invariant calculated using this gauge will give us the number of edge states on the left boundary of a right semi-infinite chain. Using either of the Bloch eigenvectors, the Zak phase index \eqref{eq:one} becomes:

\begin{align}
\vartheta =\frac{Z}{\pi} & =  -\frac{1}{2 \pi } \int_{-\pi}^{\pi} \frac{\partial \varphi (k)}{\partial_k}dk \nonumber \\
    & =- \frac{1}{2 \pi i} \int_{-\pi}^{\pi} \frac{\partial}{\partial_k} \mathrm{ln}(t_1 + t_2 e^{-i k})dk 
\end{align}

and

\begin{equation}
\vartheta =  \frac{1}{2 \pi} \int_{-\pi}^{\pi} \frac{ t_2 e^{-i k}}{t_1 + t_2 e^{-i k}} =   
\begin{cases} 1 &  | \frac{t_1}{t_2} | < 1 \\
0 &  | \frac{t_1}{t_2} | > 1 
\end{cases}
\quad \left| \frac{t_1}{t_2} \right| \neq 1 \, .
\end{equation}

We can also make use of the model's chiral symmetry, using equation \eqref{eq:winding} and calculate the winding number:

\begin{equation}
\nu = \frac{i}{2\pi } \int_{-\pi}^{\pi}  \frac{1}{t_1 + t_2 e^{-i k}}\frac{\partial (t_1 + t_2 e^{-i k})}{\partial k} dk = \vartheta.
\end{equation}
Using the Wilson line in equ. (17) also yields $\gamma=\nu=\vartheta$ for this model. As mentioned above, the SSH model has only two topological phases, 0 and 1, and its edge states can be described by both a $\mathbb{Z}_2$ and a $\mathbb{Z}$ topological invariant. However, this is not always the case, as is shown in the following extension of the SSH model. 

\subsection{Extended SSH model (E$_n$SSH)}

\begin{figure}[htbp]
    \centering
    \includegraphics[width=1\linewidth]{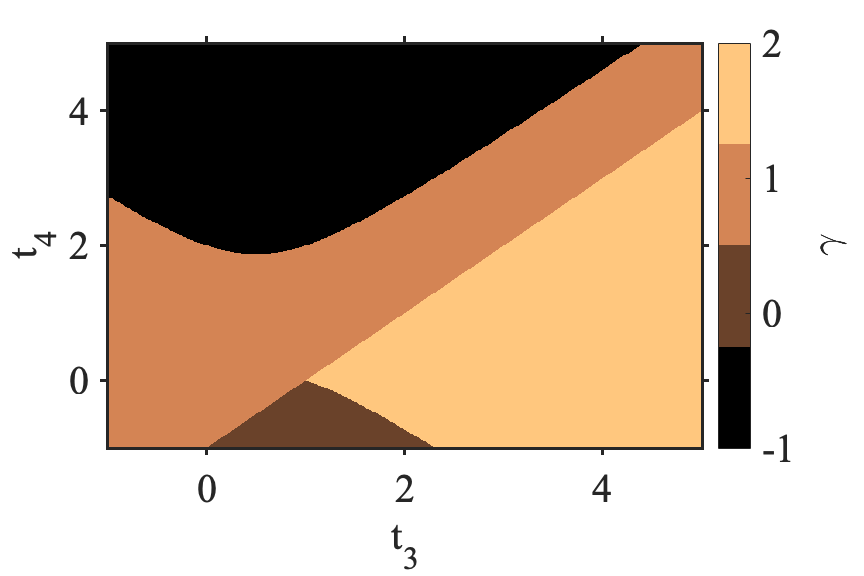}
    \caption{Wilson line invariant $\gamma$ from  eq. \eqref{eq:gamma} as a function of $t_3$ and $t_4$ with $t_1,t_2 = 1,2$ for the E$_1$SSH model.}
    \label{fig:sidebyside}
\end{figure}

\begin{figure}[h]
        \includegraphics[width=\linewidth]{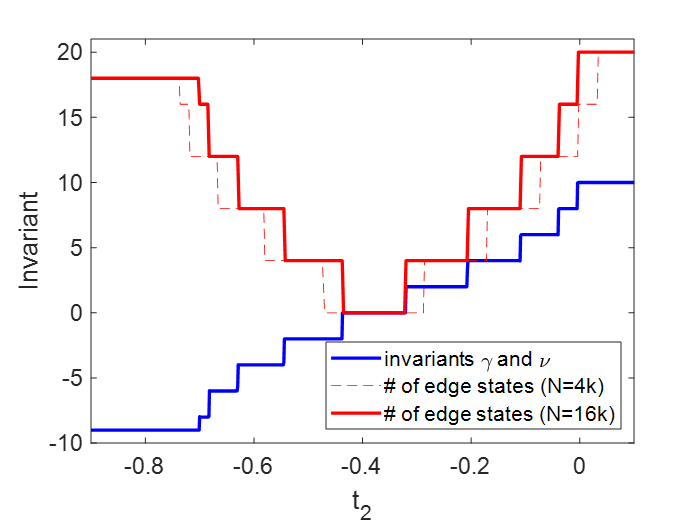}
        \caption{The winding number, $\nu$ (eq. (12)) and the Wilson line invariant, $\gamma$ (eq. (17)) of the E$_9$SSH is shown as a function of $t_2$, where ; $t_1=-t_2-0.7$, $t_3=1-t_2$, and $t_4=t_1-1$. Both, $\nu$ and $\gamma$, are computed numerically and show no visible differences. Also shown, is the number of edge states for the finite E$_9$SSH model using N=4000 and N=16000.  }
        \label{fig:LR9b}
\end{figure}

 The SSH model exhibits a much richer topological structure when long-range hopping is included (see figs. \ref{fig:sidebyside} and \ref{fig:LR9b}). Long-range hopping introduces additional couplings between non-nearest neighboring sites, altering the energy spectrum and enabling the emergence of novel topological phases \cite{perez2018ssh}. For E$_1$SSH, the winding number can take values of $\nu = -1,0,1,2$ \cite{maffei2018topological} as shown in fig. \ref{fig:sidebyside}. In this case, the $\mathbb{Z}_2$ Zak phase fails to differentiate between a topological phase of -1 and 1 or 0 and 2. 
 
In general, the Hamiltonian of this extended E$_n$SSH is as follows.

\begin{eqnarray}
H &=& \sum_{i} ( t_1 c_{A,i}^\dagger c_{B,i} + t_2 c_{B,i}^\dagger c_{A,i+1} + t_3 c_{A,i}^\dagger c_{B,i+n} \nonumber\\
&&+t_4 c_{B,i}^\dagger c_{A,i+n+1} + \text{h.c.} ),
\end{eqnarray}
where the corresponding Bloch Hamiltonian takes the form: 
\begin{equation}
H = 
\begin{pmatrix} 
0 & q(k) \\ 
q^\dagger (k) & 0 
\end{pmatrix},
\end{equation}
where $q(k) = t_1 + t_2 e^{-ik} + t_3 e^{ink} + t_4 e^{-i(n+1)k}$.

We now illustrate the Wilson line approach by numerically computing the topological invariant of the E$_9$SSH model using equ. (17). Here E$_9$SSH means $A_i\!\to\!B_{i+9}$ and $B_i\!\to\!A_{i+10}$, and all computations use $q(k)$ from equ. (24). This model exhibits a nice series of quantized invariants from -9 to 10 as shown in fig. \ref{fig:LR9b}. Note that the winding number computed numerically from equ. (12) gives identical results to $\gamma$ computed using (17), which is expected. However, in the case of a multiband system, $\nu$ and $\gamma$ give distinct results (see section C). 
In experimental systems described by SSH-like models, the topological invariant can be identified by topological edge states using the bulk-boundary correspondence. Indeed, many experimental systems are finite and possess two edges. For an even-length chain, the number of edge states comes in pairs (left and right edge states). Hence $\gamma=\frac{1}{2}\#$ of edge states. Numerically, we can evaluate the number of edge states from the eigenvectors and eigenvalues of the finite-size E$_9$SSH model. For a finite system of size $N$, the onset of edge states in terms of the hopping parameters $t_1, \cdots, t_4$ depends on the size of the system. In fig. \ref{fig:LR9b} we used sizes of $N=4000$ and $N=16000$ in order to approach the infinite limit. Note that even in the SSH model, the onset of edge states depends on the size of the system, where it is given by $t_1/t_2=N/(N+2)$ \cite{Delplace2011zak,zaimi2021detecting}. Due to the inversion symmetry of the E$_n$SSH model, the number of left- and right edge states is equal (for an even chain length). We find, as expected, that the total number of edge states is twice the absolute value of the Wilson line topological invariant, $\gamma$.

For a finite system, the exact position of the topological phase transition from one invariant to the other is given in the limit where $N\rightarrow\infty$. Numerically, the number of edge states in the E$_n$SSH model, can be obtained by looking at the density of states. The number of edge states is then simply the number of states inside the gap centered at $E=0$. As we see in fig. \ref{fig:LR9b}, only a very large $N$ gives the correct positions of the distinct topological phases.

An important property of the topological edge states, is their robustness to the presence of chiral symmetry-preserving disorder. Adding a random number $V_\text{dis}$ to each hopping term $t_i\rightarrow t_i+V_\text{dis}$ does not affect the chiral symmetry for $E=0$. Hence, zero energy edge states are preserved. For example, when looking at the topological phase $\gamma=10$ in fig. \ref{fig:LR9b}, adding $V_\text{dis}$ doesn't affect the $2\gamma=20$ edge states at $E=0$. This behavior is similar to many SSH-like models \cite{perez2019interplay}. However, the disorder cannot be too large, otherwise, the bulk gap closes.

\begin{figure}[h]
        \includegraphics[width=.9\linewidth]{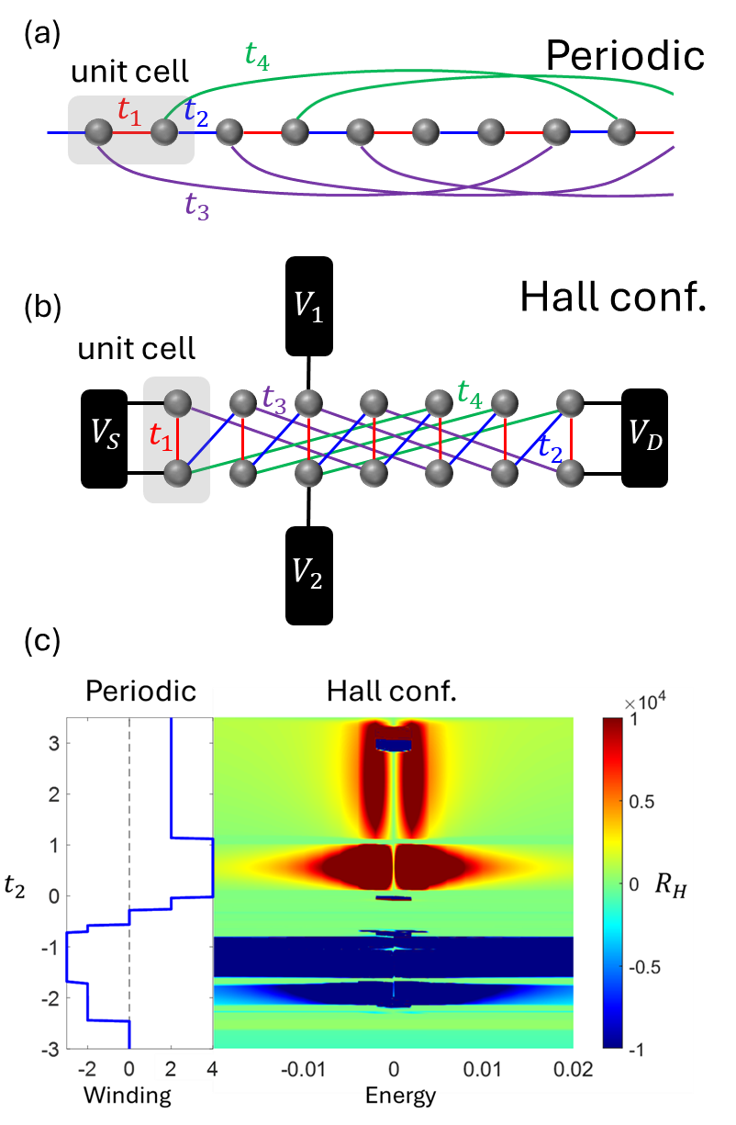}
        \caption{(a) the linear representation of the periodic E$_3$SSH model. (b) The Hall configuration of the finite  E$_3$SSH model. (c) The Wilson line invariant $\gamma$ of the periodic E$_3$SSH model as a function of $t_2$ and the Hall resistance as a function of $t_2$ and energy of the finite E$_3$SSH model for $N=800$ and all couplings to the leads are set to $t_L=0.3$. The Hall voltages ($V_1$ and $V_2$) are 3 atoms away from $V_S$ (like in the illustration).  The other hopping parameters are given by $t_1=-t_2-0.7$, $t_3=1-t_2$ and $t_4=t_1-1$.}
        \label{fig:Hall}
\end{figure}

The number of edge states only gives the absolute value of $\gamma$ (divided by two). It would be interesting to find a method, which could measure the sign change of $\gamma$ as shown in fig. \ref{fig:LR9b}. In two-dimensional systems, the toplogical Chern number, can be measured by the Hall conductance in the quantum Hall regime. In 1D models, there is no Hall conductance, but the unit cell of the extended SSH models has 2 atoms, so it is possible to measure the transverse voltage along the atom unit cell as illustrated in the fig. \ref{fig:Hall}b. Since the sign of $\gamma$ is related to the relative phase of the A and B atoms of the unit cell \cite{matveeva2023one}, we can extract the sign of $\gamma$ in an equivalent four-terminal Hall configuration. To illustrate this point, we computed the Hall resistance (transverse resistance) using a finite-sized E$_3$SSH model where the voltage contacts are weakly coupled to the chain. We have contact probes, $V_S$ and $V_D$, which are the source and drain contacts, and $V_1$ and $V_2$ for the transverse Hall contacts. We used the standard four-terminal Fisher-Lee expression to evaluate the Hall resistance \cite{fisher1981relation,buttiker1986four}. For a finite chain of length $N$ given by the Hamiltonian $H_N$, the four-terminal Hall resistance configuration in this formalism is given by $R_H=R_{2,S}-R_{1,S}$ ($S$ the source, 1,2 the voltage contacts, and $D$ the drain), where 

\begin{equation}
R=\begin{pmatrix} \sum_{j\neq S}T_{S,j}&-T_{S,1}&-T_{S,2}\\-T_{1,S}&\sum_{j\neq 1}T_{1,j}&-T_{1,2}\\
-T_{2,S}&-T_{2,1}&\sum_{j\neq 2}T_{2,j}    \end{pmatrix}^{-1}
\end{equation}
$T_{ij}=\text{Tr}(\Gamma_iG\Gamma_jG^\dagger)$, $\Gamma_i=\Sigma_i-\Sigma_i^\dagger$, and $G=(E-H_N-\sum_{i}\Sigma_i)^{-1}$. The self-energies are zero matrices of the size of $H_N$ with the non-zero diagonal element $\Sigma_{i}=|t_L|^2e^{ik}$ at the position of the leads ($i=S,1,2, \text{ or } D$), where $E=2\cos(k)$. We assumed semi-infinite leads of hopping element unity and coupled to the E$_n$SSH chain by $t_L$. Using wide band leads doesn't change the results. In Fig. \ref{fig:Hall}c we show the Hall resistance as a function of energy together with the Wilson line invariant computed using equ. \eqref{eq:gamma}. The Hall resistance is computed for a finite chain length of N=800, while $\gamma$ is obtained for the periodic system. Only in the limit $N\rightarrow\infty$ do we expect the positions of the different phases to coincide. However, even with $N=800$ we see that the agreement is already very good and the sign of the Hall resistance is negative when $\gamma$ is negative and vice versa. Note that we do not expect a quantization of the Hall resistance, unlike what is observed in two-dimensional topological insulators. Nevertheless, this Hall configuration would allow for the determination of the sign change of $\gamma$. Hence, combining this Hall configuration with a measurement of the number of edge states would fully characterize the topology of an extended SSH model. 

\subsection{SSH$_4$ Model}

The SSH$_4$ model is an extension of the SSH model to a 4-atom unit A-B-C-D, as shown in Fig. 1, resulting in two occupied bands. The Bloch Hamiltonian for the SSH$_4$ model is given by

\begin{equation}
\mathcal{H}_{\text{SSH$_{4}$}} = 
\begin{pmatrix}
0 & t_1 & 0 & t_4 e^{-i k} \\
t_1 & 0 & t_2 & 0 \\
0 & t_2 & 0 & t_3 \\
t_4 e^{i k} & 0 & t_3 & 0
\end{pmatrix},
\end{equation}

\subsubsection{Winding Number and Polarization}
The SSH$_4$ model can be mapped onto a dimer SSH$_2$ model \cite{bid2022topological,maffei2018topological} and thus only takes winding numbers of 0 and 1. The chiral symmetry of this model implies that the Hamiltonian can be written as equ. \eqref{eq:offdiag},
where \( Q(k) \) takes the following form:

\begin{equation}
Q(k) = 
\begin{pmatrix}
t_1 & t_4 e^{-i k} \\
t_2 & t_3
\end{pmatrix}.
\end{equation}

The winding number using equ. \eqref{eq:winding} is given by

\begin{equation}
\nu = \begin{cases} 
1, & \text{if } |t_1 t_3| < |t_2 t_4|; \\
0, & \text{if } |t_1 t_3| > |t_2 t_4|.
\end{cases}
\label{eq:ssh4_cases}
\end{equation}

Since the SSH$_4$ model generally lacks inversion symmetry, the polarization of individual bands is not quantized. However, the sum of the polarizations of the occupied bands remains quantized and gives the same phase diagram as eq. \eqref{eq:ssh4_cases}.

\subsubsection{Wilson Line Invariant $\gamma$}
Calculating $\gamma$ using eq. \eqref{eq:gamma}, we obtain the same results as the number of left edge states in a right semi-infinite chain \cite{lee2022winding}:

\begin{figure}[h!]
       \includegraphics[width=\linewidth]{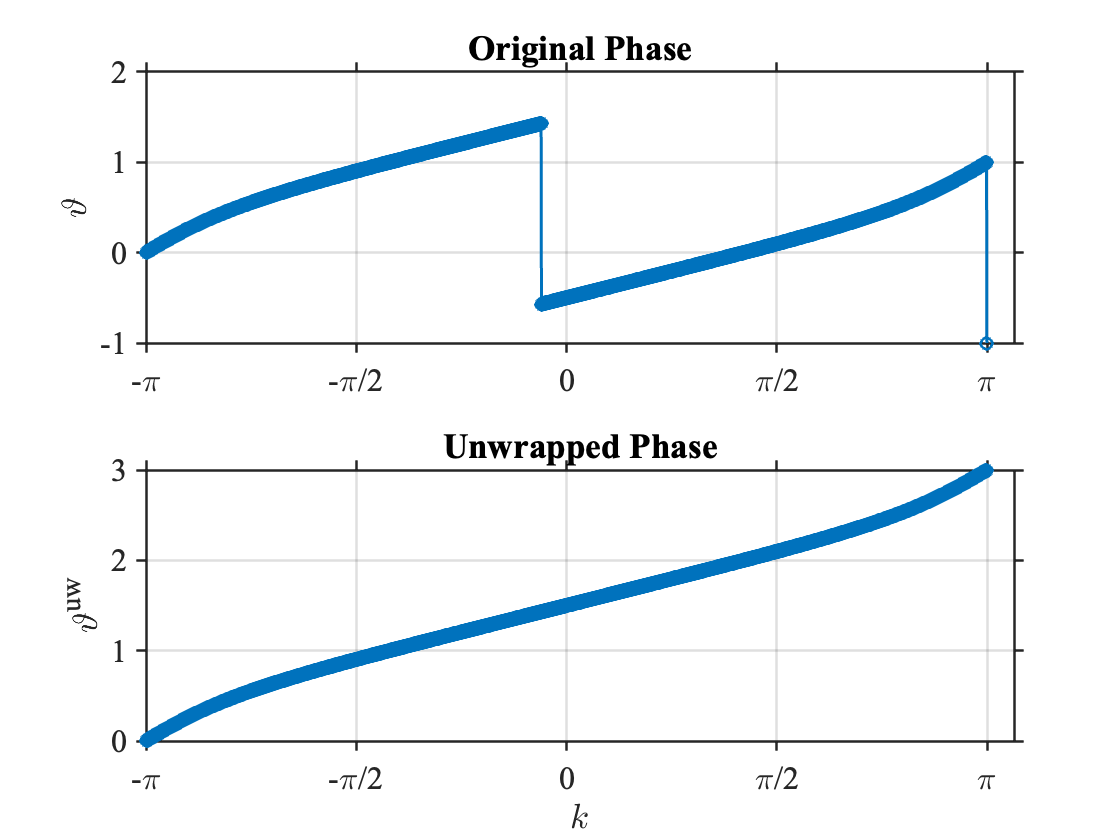}
        \caption{The top graph shows the original phase and the bottom shows the unwrapped phase for parameters $t_1=0.52,t_2=1.5,t_3=0.39, t_4=2$.
       }
        \label{fig:unwrapped}
\end{figure}

\begin{equation}\label{eq:windingresult}
\gamma = \begin{cases}
3 &  \left|\frac{t_1 t_3}{t_2 t_4}\right| < 1 \text{ and } \left|\frac{t_2 t_3}{t_1 t_4}\right| < 1 \\
2 &  \left|\frac{t_1 t_3}{t_2 t_4}\right| > 1 \text{ and } \left|\frac{t_2 t_3}{t_1 t_4}\right| < 1 \\
1 & \left|\frac{t_1 t_3}{t_2 t_4}\right| < 1 \text{ and } \left|\frac{t_2 t_3}{t_1 t_4}\right| > 1 \\
0 & \left|\frac{t_1 t_3}{t_2 t_4}\right| > 1 \text{ and } \left|\frac{t_2 t_3}{t_1 t_4}\right| > 1
\end{cases}
\end{equation}

In this case, $\gamma$ can take integer values from 0 to 3,  which predicts more edge states than indicated by polarization or winding number. Fig. \ref{fig:unwrapped} illustrates an example, where the unwrapped Wilson line yields an invariant of 3, in contrast to the polarization, which gives a value of 1.

\begin{figure}[h!]
\includegraphics[scale=0.7]{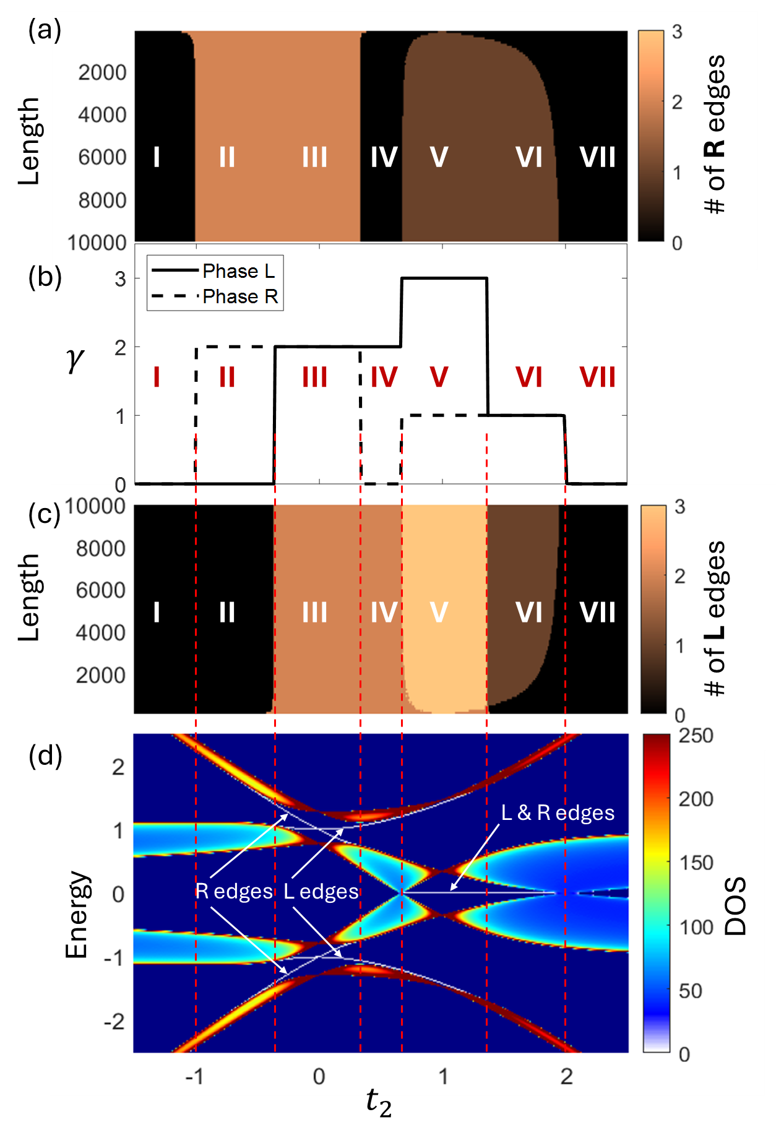}
\caption{\label{fig:SSH4} (a): Number of right edge states as a function of chain length and $t_2$. (b): The $\gamma$ invariants for phases L and R as a function of $t_2$. (c): The number of left edge states as a function of chain length and $t_2$. (d): Density of states as a function of energy and $t_2$. The hopping parameters used are $t_1=1$, $t_3=1-t_2$, and $t_4=0.5$.
}
\end{figure}

We can also compare $\gamma$ with the number of edge states by analyzing the energy spectrum of the model. These edge states appear as discrete in-gap energy levels within the bulk energy gap as shown in the density of states (DOS) in Fig. \ref{fig:SSH4}d.

For instance, in phase V (see label definition in Fig. \ref{fig:SSH4}) we can identify three edge states on the left boundary and one edge state on the right boundary. Two of the left edge states are at non-zero energies, while at $E=0$ we have both a left and a right edge state. This is in contrast to the polarization and winding number, which would detect only one edge state on each side (the $E=0$ edge states).

The results for $\gamma$ are shown in Fig. \ref{fig:SSH4}b for two possible phases of the eigenvectors. Phase L, where  $\psi_D$, the amplitude on site D is set to be real and Phase R, where $\psi_A$, the amplitude on site A is set to be real. Through the bulk-boundary correspondence, $\gamma$ for phase L will count the number of left edge states, while for phase R, it counts the number of right edge states. 

Numerically, the left and right edge states in Figs. \ref{fig:SSH4}a and \ref{fig:SSH4}c are computed by defining an edge state when the localization length of the edge state is smaller than the system length ($L_c<N$), where $L_c^{-1}$ is the average slope of $\ln(|\psi_i|)$ along the finite chain indexed by site $i$. This slope is negative for a left-edge state and positive for a right-edge state. Since the edge states are inside the gap, it is also possible to extract them by looking at the DOS. This method counts the sum of the left and right edge states. The Wilson line invariant corresponds to the topology obtained from edge states in the limit of $N\rightarrow\infty$.

When varying the hopping couplings, it is possible to go through a number of different topological phases, which we denoted by I, II, III, IV, V, VI, and VII in Fig. \ref{fig:SSH4}. Phase I is the trivial phase, where $\gamma=0$ and no edge states exist. In phase II there will be two right edge states at non-zero energies $E_\pm$ and $\gamma=2$ for phase R and $\gamma=0$ for phase L. In phase III we have two left edge states and two right edge states at $E_\pm$; $\gamma=2$ for both phases L and R. In phase IV we only have two left edge states at $E_\pm$, while in phase V we have, in addition, one left edge state and one right edge state at zero energy. In phase VI, there are only two left- and right-edge states at zero energy, and phase VII is again the trivial phase. 

\subsubsection{Robustness of Topological Edge States to Disorder}

\begin{figure}[b!]
       \includegraphics[width=0.9\linewidth]{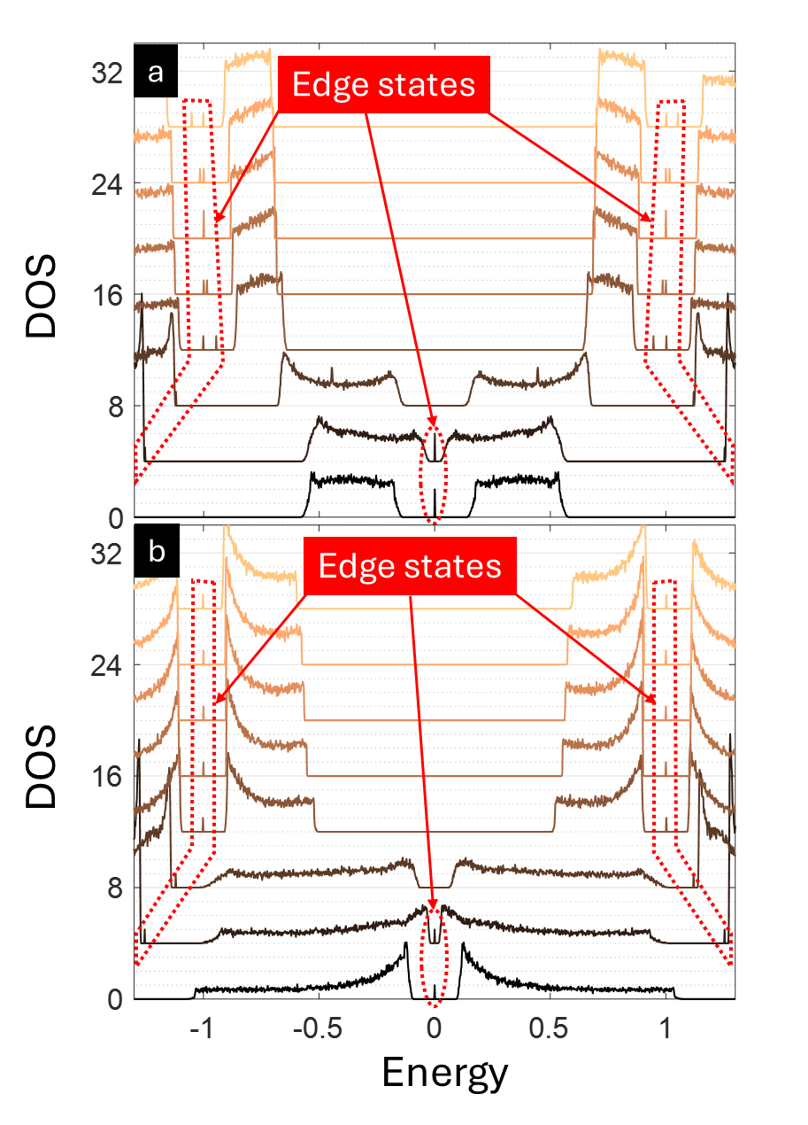}
        \caption{The figures show the DOS as a function of energy for various values of $t_2=1.105,0.755,0.505,0.055,    0.015,0,-0.015,-0.05$ from bottom to top with an offset in the DOS of 4. The other hopping terms are the same as in figure \ref{fig:SSH4}. We used a chain of length $N=2000$ and 100 averages over different disorder configurations. For the top figure we used off-diagonal disorder D-A with disorder strength $W=0.25$ and on the bottom figure onsite disorder A with disorder strength $W=1$.}
        \label{fig:disorder}
\end{figure}

To further identify the topological nature of the edge states, we can look at their behavior as a function of disorder. The $E=0$ edge states are protected by chiral symmetry preserving disorder, such as off-diagonal disorder ($t_i\rightarrow t_i+V_\text{dis}$). We characterize the strength of the disorder with $W$, where the random variable, $V_\text{dis}$, is uniformly distributed within $-W < V_\text{dis} <W$. In contrast to the $E=0$ edge states, the non-zero energy edge states ($E_\pm$) are not protected by conventional chiral symmetry but rather by a hidden chiral symmetry, dubbed $C_{1/2}$ \cite{Marques2019OnedimensionalTI}. For $C_{1/2}$ symmetry to protect the in-gap states, the SSH$_4$ chain must have the four hopping amplitudes  arranged so that the following conditions are met. First that the chain remains bipartite and secondly no cyclic shift of the hoppings makes the sequence palindromic ($t_1 \neq t_3$ and  $ t_2 \neq t_4$), so that the physical inversion axis cannot be centered in any unit cell.  

Disorder that follows the sublattice structure of the SSH$_4$ model is particularly relevant. For instance, we define dilute onsite A disorder as adding random onsite energies on each site A and equivalently for B,C, and D. Similarly, for off-diagonal disorder, where we define off-diagional A-B disorder by adding random hopping elements between sites A and B (this is equivalent to adding a random value to each $t_1$ hopping pair). The same can be done for the B-C, C-D, and D-A hopping elements. We can then determine the DOS for a finite chain length and include disorder using one of the disorder types described above. After averaging over 100 disorder configurations, the obtained DOS is shown in fig. \ref{fig:disorder}. For certain types of disorder the DOS of the edge states does not vanish, while for others it vanishes. This is easy to understand, as unprotected edge states, whose energies are inside the bulk gap, will have energies that depend on disorder, hence their DOS vanish by configurational averaging, unlike protected edge sates.  For example, for off-diagonal disorder of type D-A, all edge states survive as shown in fig. \ref{fig:disorder}. For this case we can see that the DOS is 2 at $E=0$, unlike the onsite A disorder case, where only the right edge state at $E=0$ remains unaffected by disorder, leading to a DOS of 1. The DOS is obtained by counting the number of eigenvalues within an energy interval $\Delta E=4/N$, which defines a DOS largely independent of $N$.

\begin{table}
\begin{center}
\begin{tabular}{|c|| c | c | c | c |} 
 \hline
  Disorder type& L $(E=0)$ & R $(E=0)$ & L $(E_\pm)$ & R $(E_\pm)$ \\ [0.5ex] 
 \hline\hline
 Onsite A & \ding{55} & \checkmark & \ding{55} & \checkmark \\ 
 Onsite B & \checkmark & \ding{55} & \ding{55} & \ding{55} \\
 Onsite C & \ding{55} & \checkmark & \ding{55} & \ding{55} \\
 Onsite D & \checkmark & \ding{55} & \checkmark & \ding{55} \\
Off-diag. D-A & \checkmark & \checkmark & \checkmark & \checkmark\\
Off-diag. A-B & \checkmark & \checkmark & \ding{55} & \checkmark\\
Off-diag. B-C & \checkmark & \checkmark & \ding{55} & \ding{55}\\
Off-diag. C-D & \checkmark & \checkmark & \checkmark & \ding{55}\\
 \hline

\end{tabular}
\end{center}
 \caption{Table shows the different types of disorder and the robustness to disorder for the different edge states. Checkmarks indicate edge states that are robust to the corresponding disorder type.}
\end{table}

\begin{figure}[b!]

    \includegraphics[width=0.9\linewidth]{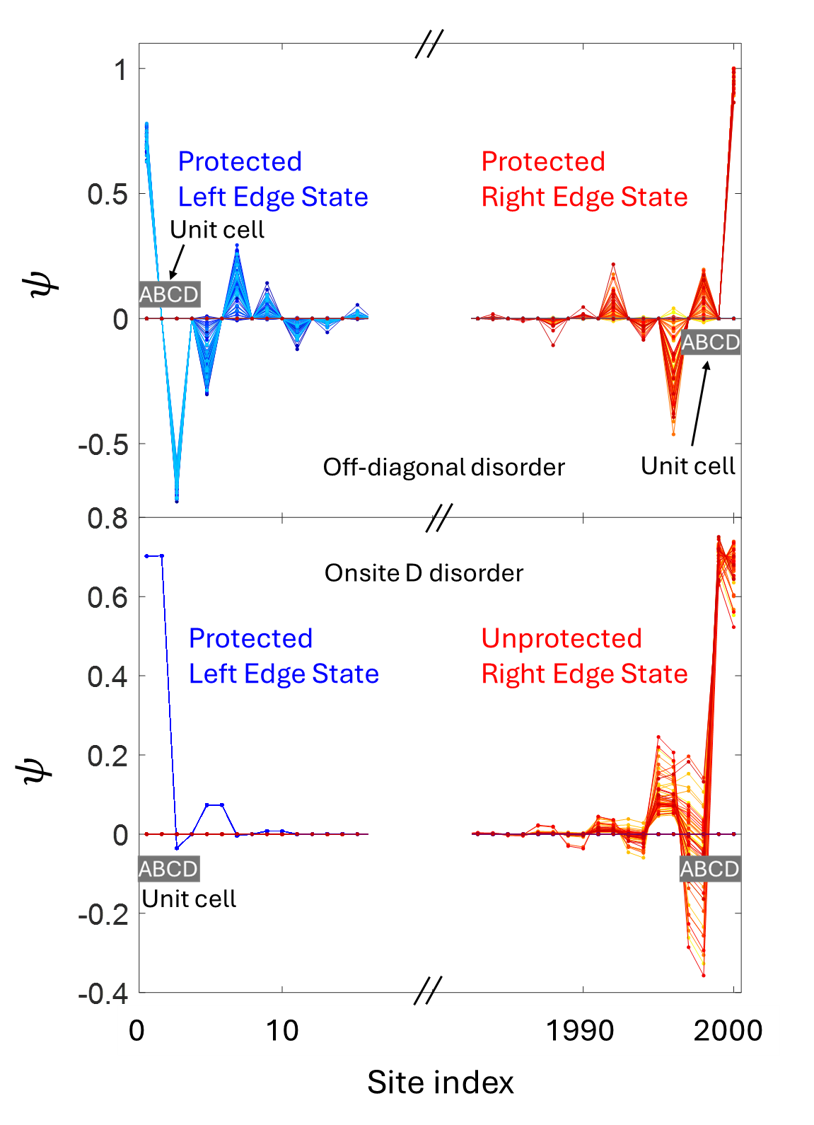} %
    \caption{wavefunction amplitude ($\psi$) distributed across the lattice sites, where we only show the 16 left-most and right-most sites of a total of 2000 sites. In the top graph we show $E=0$ edge state solutions for 100 different off-diagonal disorder configurations with a disorder strength of $W=0.25$. Here we chose a general off-diagonal disorder distributed over all the hopping elements. In the bottom graph, we used onsite D disorder with strength $W=1$ for 100 different disorder configurations for the $E=E_+$ solutions. The 100 left Edge solutions are identical (only one line is visible), unlike the 100 right edge solutions (multiple lines). We used the same parameters as in figure \ref{fig:disorder} with $t_2=1.105$ for the top graph and $t_2=-0.05$ for the bottom graph. }

    \label{fig:two_figures}
\end{figure}

Depending on the disorder class, as summarized in table 1, some of the edge states are not affected by the introduction of disorder. If we combine onsite disorder A and B, all edge states vanish in contrast to generalized off-diagonal disorder (i.e., combining D-A, A-B, B-C, and C-D) that preserve the left and right edge states at zero energy due to the chiral symmetry at $E=0$. Dilute onsite disorder is also known to preserve delocalized states in random 1D models at critical energies \cite{hilke1997localization,nguyen2022quasiresonant}. This is a property of the sublattice structure. For instance, when looking at the topological edge states in the SSH$_4$ model at non-zero energy ($E_\pm$), we see that the amplitude vanishes at every D site for the left edge states as shown in fig. \ref{fig:two_figures}. Hence, with onsite D disorder, the edge states are not affected by this type of disorder. This is different for the right edge states, where in the non-disordered case, the amplitude is zero on every A site, therefore, this edge state is affected by onsite D disorder. This would be similar for the $E=0$ edge states as summarized in table I. For off-diagonal disorder, the chiral symmetry is preserved at $E=0$, which means that the $E=0$ edge states are all preserved in the sense that the energy is not affected. However, as shown in fig. \ref{fig:two_figures}, the amplitudes vary depending on the disorder configuration  but not the parity (the amplitudes are zero on B and D sites for the left edge state and zero on the A and C sites for the right edge state).

\section{Conclusion}
We have studied the topological invariant for the E$_n$SSH and SSH$_4$ models in some detail. Our results can be generalized to other models such as SSH$_n$ as long as each band is gapped. In this case, the number of bands is given by $n$ and $n$-1 band gaps with corresponding edge states. The case of the SSH$_3$ has been extensively studied in \cite{anastasiadis2022bulk} and provides a good discussion on a gauge invariant topological invariant and protecting symmetries.  It is also possible to add long range hopping to the SSH$_n$ model, which we label E$_m$SSH$_n$, where $m$ and $n$ can be arbitrary integers. While the general solutions can quickly become quite involved for large $m$ and $n$, calculating the $\mathbb{Z}$ topological invariant using the Wilson line approach we introduced in equ. \eqref{eq:gamma} is numerically straightforward and allows one to label all the topological phases with their corresponding invariants. We should also note that the 1D models considered here are somewhat special in the sense that they are often described as boundary obstructed topological phases \cite{khalaf2021boundary} in contrast to topological phases indexed by Chern numbers. Non-zero Chern numbers do not exist in 1D.  

Extensions to higher dimensions are also possible in the context of SSH-type models \cite{obana2019topological}, which also lead to rich topological phases \cite{ma2022electronic}. Here, we restricted ourselves to extended 1D models, which also have a similarly rich topological phase diagram with multiple invariants (see for example, figs. \ref{fig:LR9b} and \ref{fig:Hall}). We provide here a clear connection between the topological invariant and the finite-size edge-state picture using the bulk-boundary correspondence. A finite-sized system, as relevant for many experimental systems, allows the identification of the topological invariant, but not the sign change. To circumvent this limitation, we propose a new configuration, which would also allow one to probe of the sign of the topological invariant in extended SSH models. We further make a clear connection between the symmetries of the edge states and their robustness to disorder. It is indeed possible to "kill" some of the edge states with disorder while preserving other ones in the same system. This extends our toolbox even further to probe and manipulate topological states. 

We acknowledge support from INTRIQ, RQMP, NSERC, and FQRNT.

\bibliography{ESSH2}

\begin{thebibliography}{40}
\bibitem [{not({\natexlab{a}})}]{note1}%
  \BibitemOpen
  \href@noop {} {}\bibinfo {note} {Similar to \cite{maffei2018topological}, if
  $Q$ is a matrix, we can write the corresponding eigenvalue equation as $Q
  K_n=q_n K_n$, with eigenvalues $q_n(k)=|q_n(k)|e^{i\omega_n(k)}$ and
  eigenvectors $K_n$. In this case, the winding number (the total phase
  accumulated around the origin) can also be written as $\nu=\frac{1}{2\pi
  }\sum_n[\omega_n(\pi)-\omega_n(-\pi)]$, which is equivalent to equ. (12).
  Here we used
  $\text{Tr}(Q^{-1}\partial_kQ)=\sum_n\partial_k\log(q_n)$.}\BibitemShut
  {Stop}%
\end{thebibliography}

\end{document}